# Low temperature carrier transport mechanism and photo-conductivity of WSe$_2$

Manjot Kaur[1], Kulwinder Singh[1], Ishant Chauhan[1], Hardilraj Singh[1], Ram K Sharma[2], Ankush Vij[3], Anup Thakur[4], Akshay Kumar[1,*]

[1]Advanced Functional Materials Laboratory, Department of Nanotechnology, Sri Guru Granth Sahib World University, Fatehgarh Sahib-140 406, Punjab, India.

[2]Centre for Interdisciplinary Research, University of Petroleum and Energy Studies (UPES), Dehradun- 248007, Uttarakhand, India.

[3]Department of Physics, University of Petroleum and Energy Studies (UPES), Dehradun- 248007, Uttarakhand, India.

[4]Department of Basic and Applied Sciences, Punjabi University, Patiala, Punjab 140 002, India.

*Corresponding author: akshaykumar.tiet@gmail.com

## Abstract

This work reports the electrical-transport and temperature-dependent photoconductivity in tungsten diselenide (WSe$_2$) thin films. The electrical conductivity analysis shows the presence of the three regions with temperature variation. At lower temperatures (< 190K), carriers become localized to small regions in the film due to the Mott's hopping mechanism. The middle region (190 – 273 K) follows Seto's parameters and obtained low barrier height (0.0873 eV) may be responsible for the improved carrier mobility. At higher temperature (> 273K) region, thermally activated conduction is dominated with two activation energies of ~138 meV and 98 meV. The peaks obtained in photoluminescent analysis attributes to the presence of mid-bandgap states or defect states which play an important role in the photoconductivity of WSe$_2$. The transient photoconductivity measurements show consistent temperature-dependent behaviour. The effect of light intensity and wavelength variation on the photoconductivity of WSe$_2$ thin films is also discussed. The photocurrent is 1.19*10$^{-5}$ A at 125 K while at 350 K it was observed to be 3.12*10$^{-4}$ A. The light-on/off current cycles show that the current can recover to its initial state which points to the stable and outstanding reversible properties of the WSe$_2$ thin film device to be used in photodetector applications.

## 1. Introduction

Transition metal dichalcogenides (TMD) materials have recently shown a new pathway to advance energy-efficient optoelectronics and nanoelectronics[1-2]. Although graphene offers remarkable high thermal conductivity and electrical mobility but gapless nature limits its use[3-4]. Tungsten diselenide (WSe$_2$) is an important TMD material that overcome the limitations of graphene owing to its band gap (~1.2−1.7 eV)[5-6]. This feature makes them desirable for low-dimensional electroncis[7]. Unlike other TMDs, opto-electronic properties of WSe$_2$ has been less studied. WSe$_2$, a layered semiconductor, exhibits p-type conduction[8]. The previous studies on the opto-electronic properties of WSe$_2$ show promising results with high photo-responsivity but most of the studies were performed at room temperature or high temperature. At lower tempaeratures, device properties change drastically due to various effects such as – an increase of contact resistance, a decrease of thermionic emission, variation of recombination centres, trap states, etc. Thus, in order to further increase the applicability of material, realization of temperature-dependent properties especially low temperature is important. It is significant to understand how material behaves in the presence of light and at low temperature. For the deposition of WSe$_2$ thin films, various methods such as electrodeposition, chemical vapor deposition, chemical bath deposition, soft selenization, solid state reaction, thermal evaporation, galvenostatic routes, spray pyrolysis, van der Waals reheotaxy, has been employed in the previous reports[9-15]. The structural and optical properties of the fabricated thin films are reliant on the method of deposition. Due to its suitability, high deposition rate, simplicity and reproducibility, the thermal evaporation method is the most common in above said deposition techniques.

In the present work, the temperature-dependent electrical conductivity of $WSe_2$ thin films has been explored. $WSe_2$ thin films were fabricated by thermal evaporation method and then examined the possibility of a different type of carrier transport mechanisms and photoconductivity in the deposited films. It is found that three mechanisms are evident in carrier transport in $WSe_2$ thin films – low temperature Mott's variable range hopping, middle temperature range Seto's grain boundary effect whereas higher temperature range dominated by thermally activated band conduction. The photoconductivity mechanism is also discussed in detail.

## 2. Experimental methods:

$WSe_2$ has been synthesized by a solvothermal approach. For the synthesis, tungsten oxide, selenium powder and acetone in appropriate amount were put in a specially designed SS-304 autoclave. The sealed autoclave was maintained at 600 °C for 24 hours in a muffle furnace. Then the powder was obtained and characterized using the X-ray diffraction (XRD) technique. Afterward, the synthesized $WSe_2$ powder was used for thin film preparation by the thermal evaporation technique on pre-cleaned glass substrates. Thin film deposition was done in a vacuum controlled chamber using Hind HIVAC system (Model: BC-300) at a deposition rate of 10 Å/s. In the deposition process, the boat was adjusted under the glass slide lid. A known amount (200 mg) of synthesized $WSe_2$ nanomaterial was put in the boat. The pressure of the chamber was maintained at $5.3 * 10^{-6}$ mbar. After deposition, the film was taken out from the chamber and characterized by various characterization techniques.

Structural analysis was done using X-ray diffraction (XRD) technique. The surface morphology and cross-section analysis of deposited thin film were studied using a scanning electron microscope (SEM). Elemental mapping was also done. To study the electrical properties, a simple device was fabricated. The temperature dependent electrical properties measurements were done using MS-TECH probe station with a Keithley Sourcemeter-2450.

The photoconductivity measurements were done using white light of different intensities and blue light sources. The light intensity is observed using a digital luxmeter. The photocurrent is calculated by subtracting the dark current ($I_D$) from the current obtained in the presence of light ($I_L$).

**3. Results and Discussion**

Fig. S1 shows the XRD spectra of the $WSe_2$ powder and film. The broad pattern in the case of $WSe_2$ film confirms the amorphous growth of film. Fig. 1 (a) shows the SEM image of the deposited $WSe_2$ film on a glass substrate which shows a uniform deposition of the film via thermal evaporation of the material. The cross-section image shows the average thickness of ~100 nm for the deposited film (as shown in Fig. 1(b)). The elemental mapping is shown in Fig. 1 (c-f) confirms the presence of tungsten and selenide in deposited films.

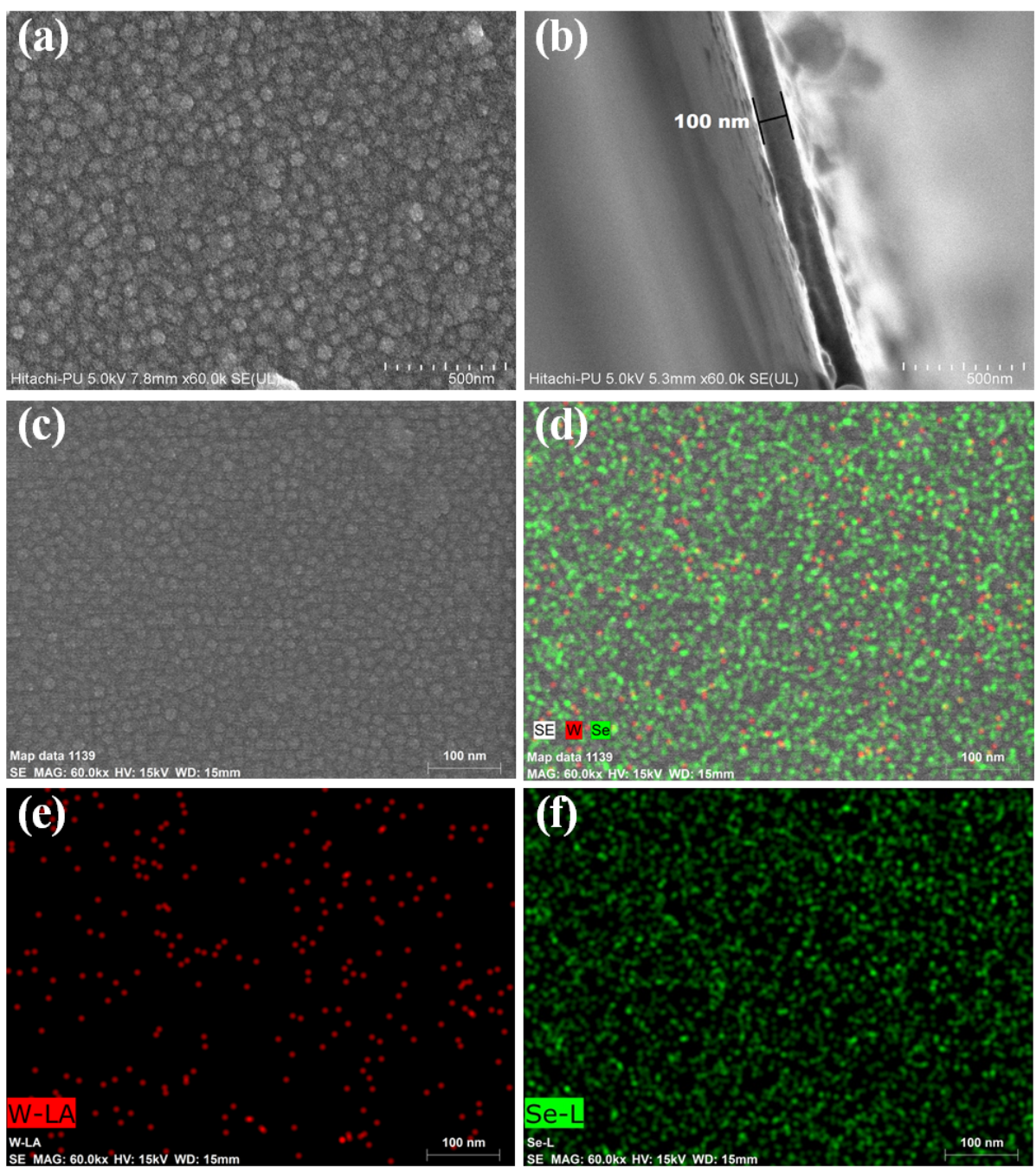

**Fig. 1:** (a) SEM image of WSe$_2$ film; (b) Cross-section image showing thickness of deposited film; (c-f) Elemental mapping of WSe$_2$ film.

Further, the temperature-dependent electrical properties has been analyzed of fabricated thin films. Fig. 2(a) shows the resistivity variation with temperature for the WSe$_2$ film in the temperature range from 125 to 348 K. There is a decrease in resistivity with the increases in temperature which shows the semiconducting behaviour of deposited WSe$_2$ film.

The three different regions in the resistivity plot with temperature indicate that there may be more than one carrier transport mechanisms involved in WSe$_2$ film. Firstly, the thermally activated band conduction mechanism has been analyzed in which, the conductivity (σ) can be expressed using Arrhenius relation;

$$\sigma = \sigma_0 \exp(-E_a/kT) \qquad (1)$$

where $\sigma_0$ is a constant, $E_a$ is the activation energy for dc conduction, and k is the Boltzmann's constant. Fig. 2(b) shows the ln σ vs 1000/T plot for WSe$_2$ film at a temperature range from 350 K to 273 K. A straight line fit to the data (region I) shows that thermal conduction dominates in the high temperature range. Two distinct thermal activation energies have been found in the plot for WSe$_2$ film whose values are calculated to be ~138 meV, 98 meV. This process includes thermal activation of carriers from donors to the conduction band.

The value of $E_a$ depends on the impurity energy levels and accepter carrier concentration. The Fermi level can be shifted up with an increase in the concentration of accepter carrier which leads to a decrease of the $E_a$. The film deposition leads to the establishment of trap states within the amorphous disordered structures. In the range of 273–125 K, carriers do not have adequate amount of energy to jump to the conduction band due to entrapment in trap states. Thus, carriers can conduct via hopping from one to another energy level in the impurity band. It signifies the domination of the hopping mechanism in a lower temperature range than thermal conduction. Hopping conduction mechanisms are of two types – Nearest Neighbor Hopping (NNH) and Variable Range Hopping (VRH)[16]. The carriers hop to the nearest neighbour empty site in NNH conduction.

For the NNH mechanism, temperature dependence can be written as[17]

$$\sigma = \sigma_{0NNH} \exp(-E_{NNH}/kT) \qquad (2)$$

Activation energy ($E_{NNH}$) is also required in this conduction which has a lower value than the activation energy ($E_a$) of thermally activated conduction. But in Fig. 2(b), ln σ vs 1000/T plot gives higher activation energy from 273 to 300 K, which means there is no dominance of NNH hopping. Hence, thermal conduction is the only possible mechanism to be responsible for conduction from 273 to 350 K with two activation energies.

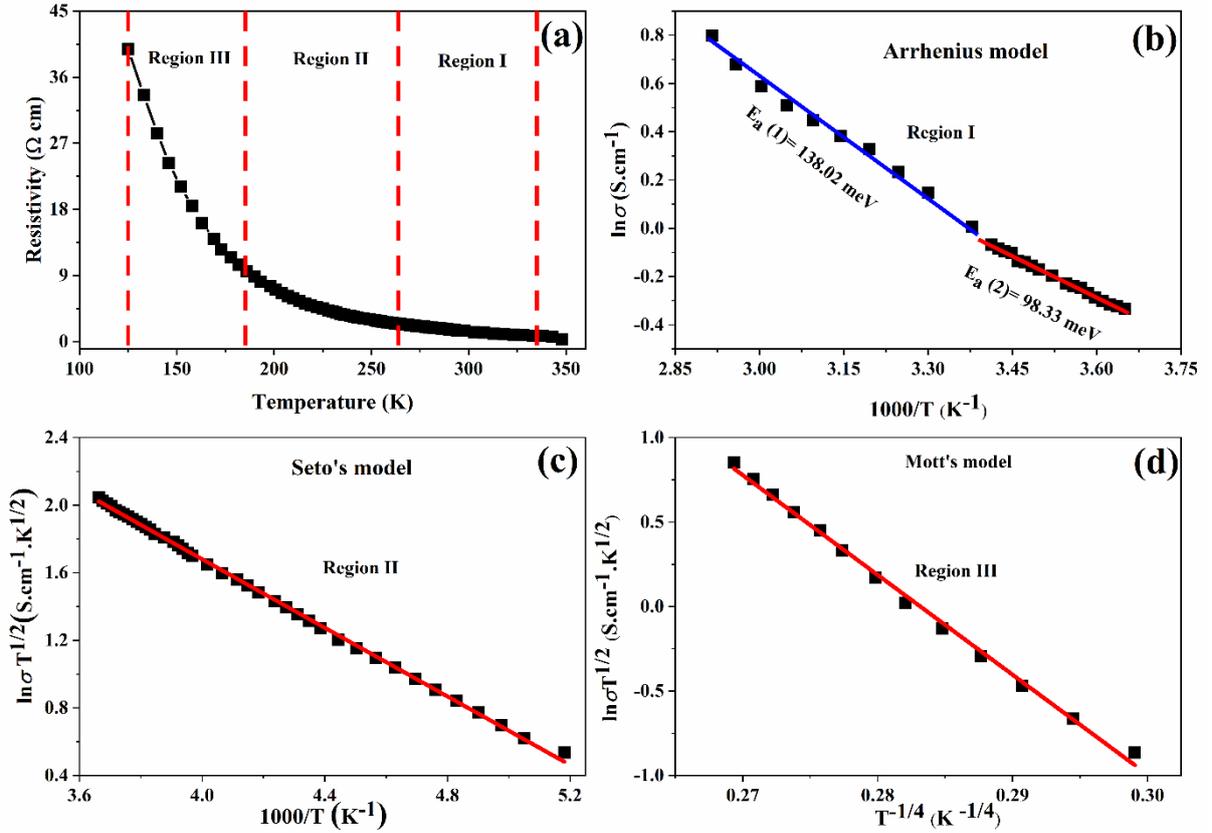

**Fig. 2:** (a) Electrical resistivity of WSe$_2$ film as a function of temperature (range 25 K to 350 K). The regions I to III have been fitted with different models; (b) Arrhenius model fits on the conductivity data in the temperature range 350 to 273 K; (c) Seto's s model fits on the conductivity data in the temperature range 273 to 190 K; (d) Mott's VRH model fits on the conductivity data in the temperature range 190 to 125 K.

We noticed the governance of Mott's VRH conduction mechanism in the lower temperature region (T<190 K). In Mott's VRH, carriers hop between the levels close to the Fermi level regardless of spatial distribution. Hopping distance is not constant in VRH because carriers hop to a site that requires the minimum possible activation energy[17]. The density of states remains the same near the Fermi level in this mechanism[18]. In Fig. 2(d), we consider Mott-VRH at a temperature range of 125 to 190 K.

In Mott's VRH, for the three-dimensional system, the temperature dependence of conductivity can be given as[18]:

$$\sigma = \sigma_{0M} T^{-1/2} \exp(-(T_M/T)^{1/4}) \qquad (3)$$

where[19]

$$\sigma_{0M} = \frac{3e^2 v_{ph}}{(8\pi)^{1/2}} [N_{(EF)}/\alpha kT]^{1/2}; \qquad (4)$$

$$T_M = [\alpha^3/kN_{(EF)}] \qquad (5)$$

$T_M$ is a measure of the degree of disorder in the film, α is the inverse of localization length, k is Boltzmann constant, $N_{(EF)}$ is the density of energy states at Fermi level, $v_{ph}$ is Debye frequency. We found that the plot of ln ($\sigma T^{1/2}$) vs. $T^{-1/4}$ fits very well the predicted Mott's 3D VRH model in the temperature range of 125 to 190 K. By best fitting of the experimental data, Mott's temperature ($T_M$) value obtained is 1217647 K. The high value of $T_M$ shows the disordered films which is also in agreement with the XRD spectrum showing amorphous nature of the deposited film. It has been found that $T_M/T \gg 1$ which follows the condition of VRH transport in semiconductor1[16, 20]. Hence, it can be assumed that VRH is the mode of carrier transport mechanism at a lower temperatures in WSe$_2$ films. Furthermore, the value of $N_{(EF)}$, mean hopping energy (W), hopping distance (R) has also been calculated by using the equations[21] given below:

$$W = 3/4\pi R^3 N_{(EF)} \qquad (6)$$

$$R = [9/8\pi\alpha kTN_{(EF)}]^{1/4} \qquad (7)$$

To satisfy Mott's VRH conduction mechanism, there are two conditions, αR>1 and W>kT[21]. The calculated values for WSe$_2$ film satisfy these conditions. So at low temperatures (<190 K), thermal energy to the carriers decreases, that's why carriers feel more interference for the electrical conduction. So, carriers become localized to small regions in the film at low temperatures.

We observed a region obtained between Mott's VRH mechanism (<190 K) and thermal conduction (>273 K), which has been explained using Seto's model. According to this model, grain boundaries act as defects which makes the crossing of boundaries difficult for carriers[22]. The linear plot between ln (σT$^{1/2}$) and 1000/T for WSe$_2$ film in the temperature range of 190 to 273 K is shown in Fig.2(c). This fitted data shows that Seto's grain boundary model applies to WSe$_2$ film in this temperature range. The fitted equation is[23]

$$\sigma = \sigma_{0S} \exp(-\Phi_B/kT) \qquad (8)$$

where,
$$\Phi_B = Le^2 n v_c / kT \qquad (9)$$

$$v_c = [kT/2\pi m^*]^{1/2} \qquad (10)$$

$\Phi_B$ is barrier height, L is grain size, e is a charge of electron, n is carrier concentration, $v_c$ is collection velocity, and m* is the effective mass of charge. The barrier height is directly proportional to grain size. From Seto's model fitting, $\Phi_B$ is 0.0873 eV which is very small showing the improved mobility of carriers. This low barrier height may be ascribed to the more conductivity of WSe$_2$ thin film. The constant $\sigma_{0S}$ has been found to be 22.58 S cm$^{-1}$. All these three temperature regions fitted by Mott's, Seto's, and Arrhenius model gives detailed information of the temperature-dependent carrier transport mechanism of WSe$_2$ thin film.

Optical studies on the WSe$_2$ films have been performed and used to calculate the band gap by Tauc's plot as shown in Fig. S2. The calculated band gap for WSe$_2$ is ~1.5 eV. Also, WSe$_2$ shows two PL peaks (as shown in Fig. S3) at ~1.71 and ~1.74 eV which can be attributed to trions and excitons respectively. These peaks are well investigated in literature[24]. Also, there is a peak at ~1.59 eV which can be ascribed to the exciton bound to defects[25,26]. This will result in mid-bandgap states above the maximum of valence band or below the conduction band minimum. These defect states play an important role in the photoconductivity of $WSe_2$ film. The photoconductivity of $WSe_2$ film is measured in the temperature range of 125-350K. Fig. 3 (a) shows the schematic of the fabricated $WSe_2$ device used for photoconductivity. Fig. 3 (b) shows the temperature dependence comparison of dark and photo conductivity. The temperature dependent I-V curves of dark and photocurrent are shown in Fig. S4 and S5. The photocurrent is calculated by subtracting dark current from current obtained in presence of light. The photocurrent is $1.19*10^{-5}$ A at 125 K while at 350 K it was observed to be $3.12*10^{-4}$ A (Fig. S6).

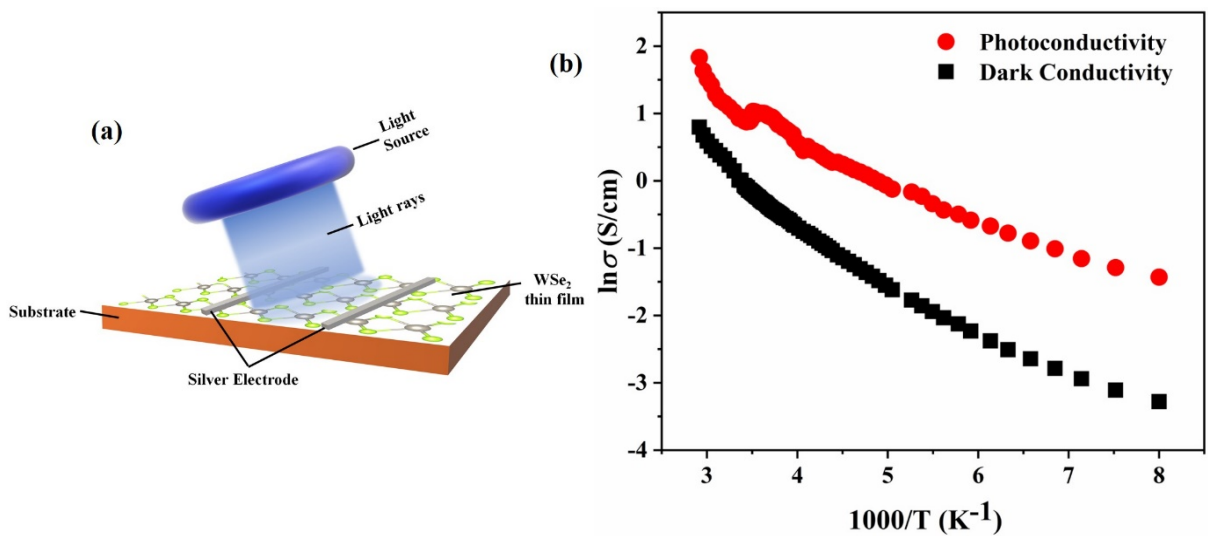

**Fig. 3:** (a) Schematic of fabricated $WSe_2$ device; (b) Variation of photoconductivity (at 1750 Lux intensity) and dark conductivity with temperature at 1750 Lux intensity;

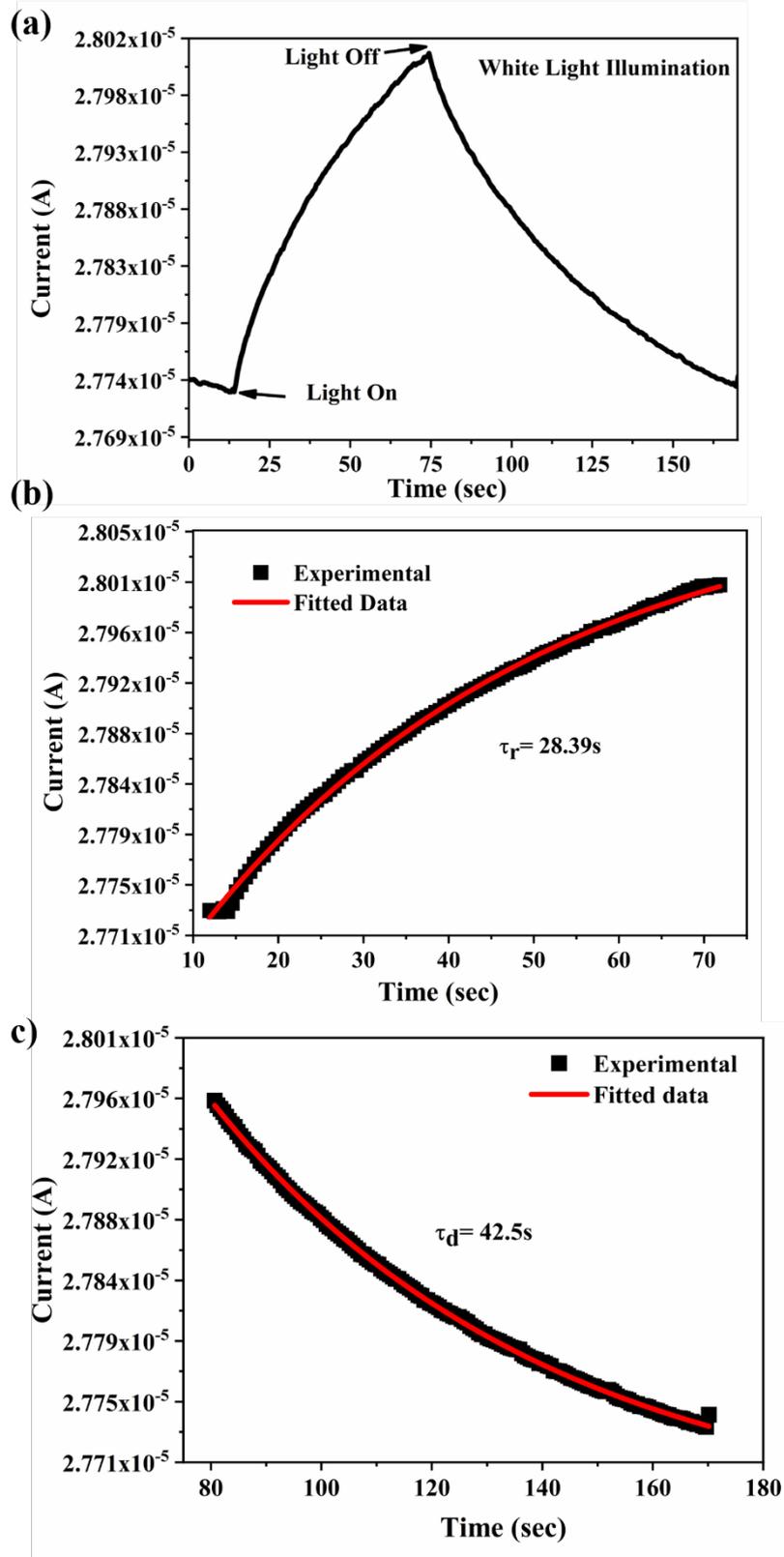

**Fig. 4:** (a) I-T curve measurement using white light illumination (1750 Lux) at room temperature; (b) Rising edge of single photo-response current cycle; (c) Decaying edge of single photo-response current cycle of WSe$_2$ film.

Fig. 4 (a) shows the single photo-response cycle of WSe$_2$ film at room temperature when illuminated with white light (Intensity = 1750 lux). The rising and decaying edge follows the equations – I(t) = I$_0$[1-exp(-t/$\tau_r$)] and I(t)= I$_0$exp(-t/$\tau_d$), where '$\tau_r$' is rising edge time constant and '$\tau_d$' is decaying edge time constant[27]. By fitting curves (Fig. 4(b) and 4(c)), $\tau_r$ and $\tau_d$ are found to be 28.39 s and 42.50 s. This shows the slow rise and decay of the photocurrent which can be found in disordered materials or due to the presence of trap states in the energy gap. According to previous reports, shallow traps generally leads to faster process while deep trap states results in a slow increase or decrease in photoconductivity[28, 29].

Also, the I-T curves of photo-response were observed by repeatedly turning on and off light illumination (white and blue both). The white light illuminated ON-OFF cycles has been shown in Fig. 5 (a). After various cycles, the current can recover to its initial state. This shows that the WSe$_2$ thin film device is stable and has outstanding reversible properties. The I-T curve with a variation of light intensity is shown in Fig. 5(b). As white light illumination power densities increased from 4.75 to 21.87 mW/cm$^2$, photo current also increased in steps. The photo-current under the illumination of different power intensities follows a power law (I ~ P$^\gamma$) with intensity exponent ɣ = 0.87 as shown in Fig. S7. Fig. 5(c) shows the photo-response curve during ON and OFF state with blue light illumination. Shallow and deep level defects in the forbidden gap influence the rising and decaying edges of photo-response. The slow decaying edge can be ascribed to the presence of trap states generally deep defects. When light is shone, photogenerated carries formed in the film. Then these carriers move from valence to conduction band. Due to the presence of deep level defects, these photogenerated carriers take a longer time to relax. Hence, defect or trap states play an important role in the photoconductivity of WSe$_2$ films.

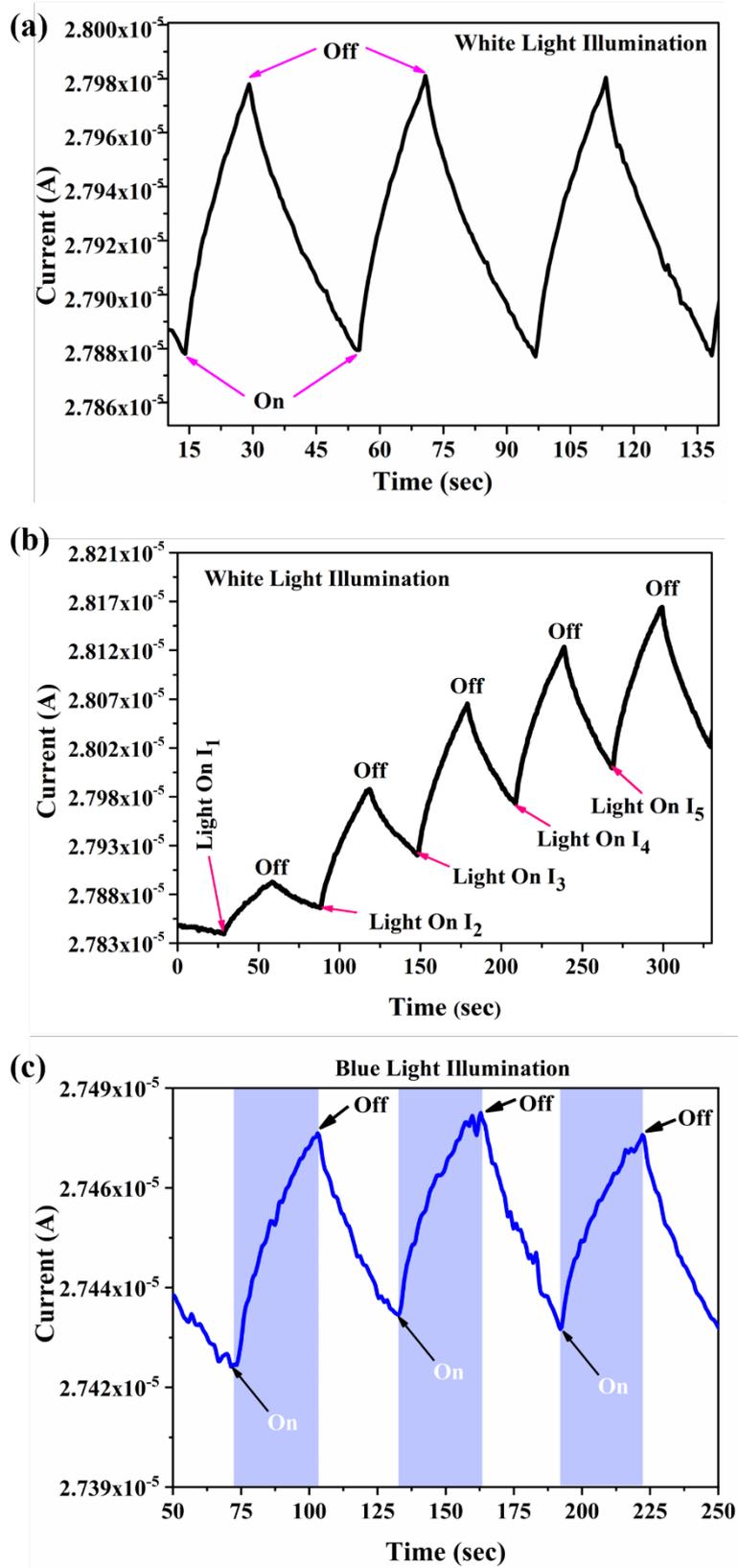

**Fig. 5:** (a) Repeated I-T curve cycles using white light illumination (1750 Lux); (b) I-T curve measurement using white light illumination of different intensities 380($I_1$), 720 ($I_2$), 1270 ($I_3$), 1580 ($I_4$), 1750 Lux ($I_5$) at room temperature; (c) I-T curve measurement using blue light illumination at room temperature.

## 4. Conclusions

This work presents the carrier transport mechanism and photo-response properties of $WSe_2$ at low temperatures. A simple thin film device with as synthesized $WSe_2$ was prepared which shows Ohmic nature. The obtained conductivity is high even at low temperatures. The carrier transport mechanism in the temperature range of 125-350 K follows three different mechanisms – Mott, Seto and Arrhenius models. Low temperature is dominated by hopping mechanism, intermediate by Seto's whereas high temperatures are dominated by thermal conduction. In Mott's mechanism, carriers become localized to small states near the Fermi level and the high value of Mott's temperature shows the disordered films. From Seto's model fitting, barrier height ($\Phi B$) has been found to be very small (0.0873 eV) showing the improved mobility of carriers. The photoconductivity also shows the same trend in different temperature regions. The obtained photocurrent follows the Power law (with ɣ = 0.87) and as white light illumination power densities increased (4.75 to 21.87 mW/cm$^2$), photo current also increased in steps. Due to the presence of deep trap states, slow rise edge (28.39 s) and decay edge time (42.50 s) has been obtained. The fabricated $WSe_2$ thin film device also displays outstanding stable and reversible photo-response properties which can be used in low temperature and high vacuum opto-electronic applications.

**Acknowledgements**

This work was supported by Council of Scientific and Industrial Research, India (SRF, file no. 09/1198(0004)/2020-EMR-I), and Science and Engineering Research Board, Department of Science and Technology under project no. EMR/2016/002815. There are no conflicts to declare.